\documentclass[aps,prd,floats,floatfix,preprintnumbers,superscriptaddress,nofootinbib]{revtex4-2}

\usepackage{graphicx} 
\usepackage{epstopdf} 
\usepackage{dcolumn}
\usepackage{amssymb}
\usepackage{amsmath}
\usepackage{bbold}
\usepackage[hidelinks]{hyperref}
\usepackage{color}
\usepackage{xcolor,colortbl}
\usepackage[utf8]{inputenc}
\usepackage[normalem]{ulem}
\usepackage{subcaption}
\usepackage{slashed}
\usepackage{verbatim}

\newcommand{\be}{\begin{equation}}
\newcommand{\ee}{\end{equation}}

\newcommand{\rmd}{\mathrm{d}}
\newcommand{\rme}{\mathrm{e}}
\newcommand{\rmi}{\mathrm{i}}

\newcommand{\delp}{\boldsymbol{\Delta}_{\perp}}
\newcommand{\xp}{\boldsymbol{x}_{\perp}}
\newcommand{\yp}{\boldsymbol{y}_{\perp}}
\newcommand{\zp}{\boldsymbol{z}_{\perp}}
\newcommand{\rp}{\boldsymbol{r}_{\perp}}
\newcommand{\ssp}{\boldsymbol{s}_{\perp}}
\newcommand{\bp}{\boldsymbol{b}_{\perp}}
\newcommand{\qp}{\boldsymbol{q}_{\perp}}
\newcommand{\qpi}{\boldsymbol{q}_{i\perp}}
\newcommand{\Pp}{\boldsymbol{P}_{\perp}}
\newcommand{\kp}{\boldsymbol{k}_{\perp}}
\newcommand{\kpi}{\boldsymbol{k}_{i\perp}}
\newcommand{\kpij}{\boldsymbol{k}_{i(j)\perp}}
\newcommand{\kpijj}{\boldsymbol{k}_{i(jj')\perp}}
\newcommand{\kpijjj}{\boldsymbol{k}_{i(jj'j'')\perp}}

\newcommand{\konp}{\boldsymbol{k}_{1\perp}}
\newcommand{\ktwp}{\boldsymbol{k}_{2\perp}}
\newcommand{\kthp}{\boldsymbol{k}_{3\perp}}
\newcommand{\qonp}{\boldsymbol{q}_{1\perp}}
\newcommand{\qtwp}{\boldsymbol{q}_{2\perp}}
\newcommand{\qthp}{\boldsymbol{q}_{3\perp}}
\newcommand{\Sp}{\boldsymbol{S}_{\perp}}
\newcommand{\calO}{\mathcal{O}}

\begin{document}
\preprint{ZTF-EP-26-05}

\title{The eikonal spin-dependent Odderon and
gluon Sivers function of a proton, and its small-$x$ evolution}

\author{Sanjin Beni\' c}
\affiliation{Department of Physics, Faculty of Science, University of Zagreb, Bijenička c.~32, 10000 Zagreb, Croatia}

\author{Adrian Dumitru}
\affiliation{Department of Natural Sciences, Baruch College, CUNY,
17 Lexington Avenue, New York, NY 10010, USA}
\affiliation{The Graduate School and University Center, The City University of New York, 365 Fifth Avenue, New York, NY 10016, USA}

\author{Florian Hechenberger}
\affiliation{Center for Nuclear Theory, Stony Brook University, Stony Brook, NY 11794, USA}

\author{Tomasz Stebel}
\affiliation{Jagiellonian University, Institute of Theoretical Physics,
Łojasiewicza 11, 30-348 Kraków, Poland}

\begin{abstract}
The matrix element in the proton of the eikonal Odderon operator, with a
helicity flip, has been shown to correspond to the dipole gluon Sivers
function. We employ a three quark light-front model of the proton
to determine the Sivers function at moderately small $x_0 \sim 0.1$ and transverse
momentum $k_\perp \lesssim 1$~GeV. The model light-cone (LC) wave function
predicts the properties of $x f_{1T}^{\perp g}(x,k_\perp)$ such as its overall magnitude, the
position of its peak in $k_\perp$, and its behavior at small $k_\perp$.
We then compute numerically the
BFKL anomalous dimension characterizing the 
power-law tail at $k_\perp \gtrsim 1.5$~GeV of the gluon
Sivers function at small (but pre-asymptotic)
LC momentum fractions $x\sim x_0 e^{-1/\alpha_s}$:
$x f_{1T}^{\perp g}(x,k_\perp) \sim k_\perp^{-3.3}$.
\end{abstract}

\maketitle

\section{Introduction}

The Sivers function describes a left-right asymmetry, relative to
the spin vector, of the transverse momentum dependent parton
distributions (TMDs) in a transversely polarized proton~\cite{Sivers:1989cc,Boussarie:2023izj}. 
Finite parton
transverse momentum allows for $\Sp \times \kp$ ``spin-orbit"
correlations in the proton. These correlations are probed even in
eikonal, high-energy scattering processes which involve transfer of transverse
momentum to the partons.

We are interested here in the (dipole) gluon Sivers function of the proton which
we denote $x f_{1T}^{\perp}(x,k_\perp)$. To date, this element of the
structure of the proton is rather poorly known~\cite{Boer:2015vso}, its measurement represents
one of the main goals of the physics program at the future
Electron-Ion Collider (EIC)~\cite{Accardi:2012qut,AbdulKhalek:2021gbh}.
In the eikonal approximation the gluon Sivers function is given by the
$C$-odd component of the dipole scattering amplitude evaluated for
a transversely polarized proton, i.e.\ the so-called spin
dependent 
Odderon~\cite{Boer:2015pni,Zhou:2013gsa,Yao:2018vcg,Boussarie:2019vmk}.
In fact, the eikonal contribution to the quark Sivers function of
the proton is also obtained from the spin-dependent Odderon~\cite{Dong:2018wsp,Kovchegov:2021iyc,Kovchegov:2022kyy}.
The general correspondence of
TMDs and matrix elements of eikonal Wilson line operators at small $x$
has been established in ref.~\cite{Dominguez:2011wm}.

In this paper we determine the eikonal spin dependent Odderon and the
associated gluon Sivers function from an effective three-quark
light-cone wave function (LCwf) of the proton, whose analytical form is given in appendix~\ref{sec:LCwf-qqq}.
In the helicity basis we can interpret the transversely polarized proton matrix element as involving helicity flips between the incoming and the outgoing proton states. Even though the eikonal color current operator $J^{+a} J^{+b} J^{+c}$, sourcing the small-$x$ gluon fields in the Odderon, conserves the 
helicities of the individual quarks, a proton helicity flip is made possible by the quark orbital angular momentum \cite{Pasquini:2008ax} which generates an overlap of proton wave functions differing by one unit of orbital angular momentum.
The possibility of a proton helicity flip in scattering with Odderon exchange
has been realized a long time ago~\cite{Ryskin:1987ya,Zakharov:1989bh} albeit
the corresponding matrix element vanishes in the non-relativistic quark model. Also,
the connection to the Sivers function
is more recent~\cite{Boer:2015pni,Zhou:2013gsa,Yao:2018vcg,Boussarie:2019vmk}.
%} 

Our empirical knowledge 
of the proton suggests that %\sout{this}
the three quark
%}
LCwf should
provide an adequate model for the partonic structure of the proton for
moderately small $x\sim 0.1$, and transverse momenta up to
$k_\perp \sim 1$~GeV. To assess the accuracy of this approach,
we confront the Dirac and Pauli electromagnetic form factors
obtained in this model with up to date data in appendix~\ref{sec:Dirac-Pauli-FF}.
We then employ the eikonal Odderon derived from the non-perturbative
LC quark model
wave function as initial condition for QCD evolution to smaller $x$.
That is, we
solve the small-$x$ dipole evolution equation~\cite{Mueller:1993rr}
from this initial condition to determine the gluon Sivers function
and the anomalous dimension associated with its
transverse momentum dependence.

So far, the only model for the eikonal spin-dependent Odderon available
in the literature was constructed by Zhou~\cite{Zhou:2013gsa} within the
framework of an extended McLerran-Venugopalan model with cubic Casimir~\cite{Jeon:2005cf}.
In that model the random static sources for the soft gluon fields are assumed
to reside
in a high-dimensional representation of color-SU(3), and therefore also in
a high-dimensional representation of spin-SU(2).
Here, we construct a model where these sources are in the fundamental
representations of both color-SU(3) and spin-SU(2).

The spin dependent Odderon features in a variety of processes 
in DIS at high energies such as
the spin dependent part of the cross section for inclusive open charm production
on transversely polarized protons~\cite{Yao:2018vcg,Zheng:2018ssm}. 
Furthermore, in exclusive forward
production of $C$-even mesons \cite{Boussarie:2019vmk,Benic:2024fbf} (see also \cite{Ma:2003py}).
Other observables for the spin dependent Odderon include a
$T$-odd nucleon energy correlator (NEC) which generates a single transverse spin 
asymmetry for the energy pattern in the target fragmentation region~\cite{Mantysaari:2025mht};
as well as transverse energy-energy correlations
(TEEC) at small $x$ in DIS~\cite{Bhattacharya:2025bqa}.
Most recently, Hatta and Teryaev discussed a left-right asymmetry w.r.t.\
the spin of the {\em recoil} baryon in DIS~\cite{Hatta:2026iry}.

The physics of the spin dependent Odderon and of the gluon Sivers function
of a transversely polarized proton should not be confused with
the physics of longitudinal spin. The latter pertains, for example, to the
determination of helicity and orbital angular momentum PDFs such as $\Delta G(x,Q^2)$ and $L_G(x,Q^2)$. Indeed, these
quantities, at small $x$, involve completely different (sub-eikonal)
operators than the spin dependent Odderon considered here;
see sec.~3.3 of ref.~\cite{Proceedings:2026xrb} for a recent overview.

from the Melosh rotation $\sim\boldsymbol{\sigma}_\perp\times\boldsymbol{k}_\perp$, where $\boldsymbol{\sigma}_\perp = (\sigma^1,\sigma^2)$ are the transverse Pauli matrices, cf.\ ref.~\cite{Pasquini:2010af}. This is a purely kinematical effect, connecting the quark spin to the orbital motion in the relativistic formulation and unrelated to the conventional spin-orbit coupling $L^z S^z$ which is of dynamical origin.  The proton wave
\section{The spin dependent Odderon(s)}

The ${\cal S}$-matrix for eikonal scattering of a $q-\bar{q}$ dipole from a target proton
can be written, in dipole factorization, in the form~\cite{Iancu:2003xm,Kovchegov:2012mbw}
\begin{equation}
   1-{\cal P}_{\Lambda'\Lambda}({\boldsymbol{r}_{\perp}},{\boldsymbol{b}_{\perp}}) 
   - i{\cal O}_{\Lambda'\Lambda}({\boldsymbol{r}_{\perp}},{\boldsymbol{b}_{\perp}})
   =
\frac{1}{N_c}\frac{\langle P' \Lambda'|\, {\rm tr}\,
    V({\boldsymbol{x}_{\perp}})V^\dag({\boldsymbol{y}_{\perp}})
    |P\Lambda\rangle}{\langle P\Lambda|P\Lambda\rangle} \,.
\label{eq:dipole}
\end{equation}
Here, ${\boldsymbol{x}_{\perp}} = {\boldsymbol{b}_{\perp}} + {\boldsymbol{r}_{\perp}}/2$
and
${\boldsymbol{y}_{\perp}} = {\boldsymbol{b}_{\perp}} - {\boldsymbol{r}_{\perp}}/2$
are the transverse positions of the quark and the anti-quark, respectively, and $V({\boldsymbol{x}_{\perp}})$
are eikonal Wilson lines at transverse position ${\boldsymbol{x}_{\perp}}$:
\begin{equation}
V({\boldsymbol{x}_{\perp}}) = \hat{P}\, \exp\left(
-ig \int \rmd x^- \, A^{+a}(x^-,\boldsymbol{x}_{\perp})\, t^a\right)~.
\end{equation}
This fundamental Wilson line describes the eikonal propagation of a
quark and corresponds to the covariant derivative $(D^\mu)_{ij} = \partial^\mu
\delta_{ij} + ig A^{\mu a}(t^a)_{ij}$.

In eq.~(\ref{eq:dipole}), $\Lambda, \Lambda' =\pm1$ denote (twice) the helicities
of the incoming and scattered proton, respectively.
$C$-even Pomeron exchange relates to
${\cal P}_{\Lambda'\Lambda}({\boldsymbol{x}_{\perp}},{\boldsymbol{y}_{\perp}})$,
and the amplitude for $\Lambda'\ne \Lambda$ has been derived in ref.~\cite{Benic:2025ral}; this amplitude vanishes when the transverse momentum
transfer $\Pp'-\Pp\to 0$.  $C$-odd Odderon exchange is related to $i{\cal
  O}_{\Lambda'\Lambda}({\boldsymbol{x}_{\perp}},{\boldsymbol{y}_{\perp}})$.
A key point is that despite taking the eikonal limit, the helicity flip Odderon exchange
amplitude is non-zero even in the forward limit,
where it is integrated over the impact parameter
${\boldsymbol{b}_{\perp}}$~\cite{Zhou:2013gsa,Boussarie:2019vmk,Benic:2024fbf}.
We shall confirm this by an explicit computation of the above matrix element
at third order in $gA^+$ for a
light-cone quark model of the proton $|P\Lambda\rangle$.
Non-zero contributions to $\int\rmd^2\bp \, \rmi{\cal O}_{\Lambda'\ne\Lambda}({\boldsymbol{r}_{\perp}},{\boldsymbol{b}_{\perp}})$ arise
when the three eikonal, $t$-channel gluons attach to three {\it different}
quarks in the proton.
\\

Expanding the Wilson lines to cubic order in the field $gA^{+a}$ we
can write the Odderon amplitude in the form~\cite{Dumitru:2018vpr}
\be
\begin{split}
{\cal O}_{\Lambda'\Lambda}(\rp,\bp) & = \frac{-\rmi g^6(N_c^2 -4)(N_c^2 - 1)}{16 N_c^2} \int_{\qonp\qtwp\qthp} \frac{G_{3\Lambda'\Lambda}(\qonp,\qtwp,\qthp)}{\qonp^2 \qtwp^2 \qthp^2}\, \rme^{\rmi (\qonp + \qtwp + \qthp)\cdot\bp}\\
&\times\left[\sin\left(\frac{1}{2}\rp\cdot(\qonp - \qtwp - \qthp)\right) +
\frac{1}{3}
\sin\left(\frac{1}{2}\rp\cdot(\qonp + \qtwp + \qthp)\right)\right]\,.
\end{split}
\label{eq:oddpos}
\ee
$G_{3\Lambda'\Lambda}(\qonp,\qtwp,\qthp)$ represents the $C$-odd part of the
eikonal coupling of three static $t$-channel gluons to the proton; 
we shall provide explicit expressions below. 
We abbreviate transverse momentum integrals as $\int_{\qp} \equiv \int \rmd^2 \qp/(2\pi)^2$, and position space integrals as $\int_{\xp} \equiv \int\rmd^2 \xp$.

Performing a Fourier transform
from impact parameter to momentum transfer we obtain
the Odderon in mixed representation,
\be
\begin{split}
{\cal O}_{\Lambda'\Lambda}(\rp,\delp) &= \int_{\bp} \rme^{-\rmi \delp\cdot\bp}
{\cal O}_{\Lambda'\Lambda}(\rp,\bp) \\
 &= \frac{-\rmi g^6(N_c^2 -4)(N_c^2 - 1)}{48 N_c^2} \int_{\qonp\qtwp} \frac{G_{3\Lambda'\Lambda}(\qonp,\qtwp,\delp - \qonp - \qtwp)}{\qonp^2 \qtwp^2 (\delp - \qonp - \qtwp)^2}\\
&\hspace{-2cm}\times\left[
\sin\left(\frac{1}{2}\rp\cdot(2\qonp - \delp)\right)
+ \sin\left(\frac{1}{2}\rp\cdot(2\qtwp - \delp)\right)
+ \sin\left(\frac{1}{2}\rp\cdot(\delp-2\qonp-2\qtwp)\right)
+ \sin\left(\frac{1}{2}\rp\cdot\delp\right)\right]\,.
\end{split}
\label{eq:oddmixed}
\ee
Forward scattering of a very small dipole scales like $\sim r^3_\perp$ as the
coupling of each one of the exchanged gluons to the almost point-like
dipole operator brings about one factor of $r_\perp$.

A further Fourier transform from dipole vector $\rp$ to transverse momentum $\kp$
gives
\be
\begin{split}
\calO_{\Lambda'\Lambda}(\kp,\delp) & = -\frac{g^6(N_c^2 -4)(N_c^2 - 1)}{32 N_c^2} 
(2\pi)^2
\int_{\qonp\qtwp} \frac{G_{3\Lambda'\Lambda}(\qonp,\qtwp,\delp - \qonp - \qtwp)}{\qonp^2 \qtwp^2 (\delp - \qonp - \qtwp)^2}
\Bigg[\delta^{(2)}\left(\qonp - \frac{1}{2}\delp - \kp\right)\\
& - \delta^{(2)}\left(\qonp - \frac{1}{2}\delp + \kp\right) + \frac{1}{3}\delta^{(2)}\left(\kp - \frac{1}{2}\delp\right) - 
\frac{1}{3}\delta^{(2)}\left(\kp + \frac{1}{2}\delp\right)\Bigg]\,.
\end{split}
\label{eq:O(kT)}
\ee
Note that in this expression the anti-symmetry of the Odderon under $\kp \to - \kp$ is manifest. Hence, the $\kp^2 \rmd^2\kp$ integral 
of this function is zero; it 
would represent the imaginary part of a twist-2 PDF which, however, must be
real.

As is customary in the literature, when referring to the forward matrix element at
zero momentum transfer we will also use the notation ${\cal O}_{\Lambda'\Lambda}(\rp)
\equiv {\cal O}_{\Lambda'\Lambda}(\rp,\delp=0)$, or
${\cal O}_{\Lambda'\Lambda}(\kp)$ for its Fourier transform to transverse momentum space. We have
%When multiplied by a power $p>0$ of $k_\perp$, we have
\be
\begin{split}
\calO_{\Lambda'\Lambda}(\kp) 
&= -\frac{g^6(N_c^2 -4)(N_c^2 - 1)}{32 N_c^2} 
(2\pi)^2
\int_{\qp} \sum_{s=\pm1}\frac{sG_{3\Lambda'\Lambda}(s\kp,\qp,- s\kp - \qp)}
{\kp^2 \qp^2 ( s\kp + \qp)^2} \\
&= -\frac{g^6(N_c^2 -4)(N_c^2 - 1)}{16 N_c^2} 
(2\pi)^2
\int_{\qp} \frac{G_{3\Lambda'\Lambda}(\kp,\qp,- \kp - \qp)}
{\kp^2 \qp^2 (\kp + \qp)^2}
\,.
\end{split}
\label{eq:O(kT)-forward}
\ee
In the second step we used the anti-symmetry of $G_{3\Lambda'\Lambda}$ under
a sign flip of all its arguments, which applies when $\Lambda'=-\Lambda$,
see the following sec.~\ref{sec:eikonal-ggg-G3}.
\\

The most general angular decomposition of the Odderon in the helicity basis is \cite{Boussarie:2019vmk}
\be
{\cal O}_{\Lambda'\Lambda}(\rp,\delp) = \delta_{\Lambda\Lambda'}\rmi\cos(\phi_{r\Delta}){\cal O}(\rp,\delp) + \delta_{\Lambda,-\Lambda'}\rmi\Lambda\rme^{\rmi\Lambda\phi_r}{\cal O}_S(\rp,\delp) + \delta_{\Lambda,-\Lambda'}\rmi\Lambda\rme^{\rmi\Lambda\phi_\Delta}\cos(\phi_{r\Delta}){\cal O}^\perp_S(\rp,\delp)\,.
\label{eq:Odecompose}
\ee
There are three independent real functions in general. The first one is the usual
spin-independent (also known as ``b-dependent") Odderon while the second and third ones
are spin-dependent Odderons. The second one becomes the Sivers function in the forward limit, and that is our focus here.
%; it can be extracted from eq.~(\ref{eq:oddmixed}) via
%\begin{equation}
%    \rmi\Lambda\rme^{\rmi\Lambda\phi_r}{\cal O}_S(\rp,\delp\to 0) =
%    {\cal O}_{\Lambda'=-\Lambda}(\rp,\delp\to 0) ~.
%\end{equation}

We may also consider the matrix element of the Odderon operator in a (transverse)
spin basis which is usually employed to relate it to the gluon Sivers function.
To obtain a state of definite spin vector one constructs a superposition of
states of definite LC helicity~\cite{Diehl:2005jf}
\be
| \vec S \rangle = \cos\frac{\theta_S}{2}\, |\Lambda=+\rangle
\,+\,  \sin\frac{\theta_S}{2} \, \rme^{\rmi\phi_S}\, |\Lambda=-\rangle~.
\ee
Here, $\vec S = (\sin\theta_S \cos\phi_S, \sin\theta_S \sin\phi_S, \cos\theta_S)$.
In case of a transverse spin vector $\Sp$ we have $\theta_S=\frac{\pi}{2}$ and
\be
| \Sp \rangle = \frac{1}{\sqrt 2}\left( |\Lambda=+\rangle
\, +\,  \rme^{\rmi\phi_S}\, |\Lambda=-\rangle \right)~.
\ee
Then,
\begin{equation}
\begin{split}
    {\cal O}_{\Sp\Sp}(\rp,\delp) &= 
    \sum_{\Lambda',\Lambda}\langle \Sp | \Lambda'\rangle\, 
    {\cal O}_{\Lambda'\Lambda} \, \langle\Lambda | \Sp\rangle\\ 
    &= \cos(\phi_{r\Delta})\rmi\calO(\rp,\delp) - \frac{\Sp\times\rp}{r_\perp}\calO_S(\rp,\delp) - \frac{\Sp\times\delp}{\Delta_\perp}\cos(\phi_{r\Delta})\calO_S^\perp(\rp,\delp)\,,
\end{split}
\end{equation}
with $\Sp \times\rp = r_\perp \sin(\phi_r - \phi_S)$. In the $\Delta_\perp \to 0$
forward limit,
\begin{equation}
    {\cal O}_{\Sp\Sp}(\rp) \to - \frac{\Sp\times\rp}{r_\perp}
    \calO_S(\rp,\delp = 0)~.
\end{equation}

The gluon Sivers function of the eikonal dipole model can be obtained from its
forward, helicity flip Odderon~\cite{Boer:2015pni,Zhou:2013gsa,Yao:2018vcg}:
\be %\label{eq:f_Siv-O_S}
x f_{1T}^\perp (x,k_\perp) =
\frac{N_c k_\perp M_p}{2 \pi^3 \, g^2}
\rme^{-\rmi\phi_k} \, \calO_{-+}(\kp)\,.
\label{eq:sivodd}
\ee
On the l.h.s.\ of this expression, the Sivers function does not depend on
the direction of $\kp$.
$M_p=0.938$ GeV denotes the mass of the proton.

In Zhou's model~\cite{Zhou:2013gsa} the integral of
$x f_{1T}^\perp (x,k_\perp)$ over $\rmd k_\perp^2$ vanishes,
\be
\int\limits_0^\infty \rmd k_\perp \, k_\perp \, x f_{1T}^\perp (x,k_\perp) = 0~.
\label{eq:f-Siv-sum-rule}
\ee
Given the expressions for $x f_{1T}^\perp (x,k_\perp)$ obtained here,
eq.~(\ref{eq:f-Siv-sum-rule}) does not appear to represent an exact ``sum rule".
However, a numerical
evaluation confirms it at a level better than 0.5\%.

The above relation implies that the Sivers function is not sign-definite,
and one may expect substantial cancellations, in particular at
relatively low scales $\mu^2$, also in the
$k_\perp$-integral for the twist-3 collinear function
(also called the $d$-type tri-gluon PDF) given 
by~\cite{Kang:2008qh, Zhou:2013gsa,Boer:2015pni,Benic:2024fbf}
\be\label{eq:f-Siv-kt2-moment}
x f_{1T}^{\perp(1)}(x,\mu^2) = 8\pi\, O(x,\mu^2) = \pi
\int^{\mu^2}_0 \rmd k_\perp^2 \, \frac{k^2_\perp}{2M_p^2}\, x f_{1T}^\perp (x,k_\perp)~.
\ee
We shall investigate the magnitude of this function via
a specific physical process in sec.~\ref{sec:Results-Discussion} below.

\subsection{Eikonal triple gluon coupling to the proton}
\label{sec:eikonal-ggg-G3}

The $C$-odd coupling
of three small-$x$, $t$-channel gluons (in covariant gauge) to the proton is given
by the expectation value of three light-front color charge
density operators $\rho^a(\qp)\equiv J^{+a}(\qp)$, i.e.\ of the fundamental cubic Casimir of SU(3). These color charges source the soft gluon fields, $A^{+a}$, appearing in the
Wilson lines in eq.~(\ref{eq:dipole}).
\begin{equation}
\begin{split}
    d^{abc}
\langle P' \Lambda' |\rho^a(\qonp)\rho^b(\qtwp)\rho^c(\qthp)| P\Lambda\rangle 
&\equiv 
\frac{(N_c^2-1)(N_c^2-4)}{4N_c} \, G_{3\Lambda' \Lambda}(\qonp,\qtwp,\qthp)\\
&= N_c C_{3F} \, G_{3\Lambda' \Lambda}(\qonp,\qtwp,\qthp)
\,.
\label{eq:rho3}
\end{split}
\end{equation}
(We have stripped the r.h.s.\ of this expression of the delta-functions \cite{Dumitru:2019qec} for
conservation of LC and transverse momentum.)

To proceed, we adopt an effective light-cone three-quark model of the proton
intended to describe its partonic structure for LC momentum fractions
of roughly $10^{-1}$ (and higher), and for transverse momenta up to approximately 1~GeV.
We write the proton state in the light-front (LF) quark model in the form
\be
\begin{split}
  |P,\Lambda\rangle &= \int[\rmd x_i]\int\left[\rmd^2\kpi\right]
  \sum_{j_1, j_2, j_3} \frac{\epsilon^{j_1 j_2 j_3}}{\sqrt{N_c!}}
  \, \sum_{\{\lambda_i\}} \,
  \Psi_\Lambda(x_i,\kpi,\lambda_i)\,\, |\{x_i P^+,\kpi+x_i\Pp,\lambda_i,j_i\}\rangle\,.
\end{split}
\label{eq:|P>}
\ee
Here, $x_i$ denotes the LC momentum fraction of the $i^\mathrm{th}$ quark,
where $i=1\dots N_c=3$;
$\kpi$ is its transverse momentum relative to the center of mass transverse
momentum $\Pp$ of the proton; $j_i$ refers to its color; $\lambda_i$ to its
helicity; a flavor quantum number is not required for the present
purposes. The explicit form of the
momentum space integrals and of the three-quark LC wave function
are provided in appendix~\ref{sec:LCwf-qqq}. In appendix~\ref{sec:Dirac-Pauli-FF}
we confront the model with up to date data for the Dirac and Pauli
electromagnetic form factors of the proton.

Given a light-front Fock space expansion of the proton, 
this matrix element can be computed explicitly, in principle.
For a proton effectively
made of three valence quarks, as also assumed here,
such computations have previously been carried out in ref.~\cite{Dumitru:2018vpr,Dumitru:2019qec,Dumitru:2020fdh}, however, without
considering a helicity flip of the proton.
An analogous matrix element for eikonal {\em two}-gluon exchange {\em with}
helicity flip has been presented in ref.~\cite{Benic:2025ral}.
We refer to those papers for technical details. Here we just quote our
result,
\be
\label{eq:g3}
\begin{split}
G_{3\Lambda'\Lambda}(\qonp,\qtwp,\qthp) & = 
\sum_{\lambda_i}\sum_{j = 1}^3\int_{x_i} 4\pi \delta(1- x_1 - x_2 - x_3) \int_{\kpi}(2\pi)^2 \delta^{(2)}(\konp + \ktwp + \kthp)\\
&\times\Big[\Psi^*_{\Lambda'}(x_i, \kpij,\lambda_i) -\frac{1}{2}\sum_{l = 1}^3\sum_{j' \neq j} \Psi^*_{\Lambda'}(x_i, \kpijj^{(l)},\lambda_i)\\
& + \sum_{j' \neq j}\sum_{j'' \neq j' \neq j} \Psi^*_{\Lambda'}(x_i, \kpijjj,\lambda_i)\Big]\Psi_{\Lambda}(x_i, \kpi,\lambda_i)\,,
\end{split}
\ee
where
\be
\begin{split}
\kpij &= \kpi + (x_i - \delta_{ij})(\qonp + \qtwp + \qthp)\,,\\
\kpijj^{(l)} &= \kpi + (x_i- \delta_{ij}) (\qonp + \qtwp + \qthp) - 
(\delta_{ij'}- \delta_{ij})\boldsymbol{q}_{l\perp}\,,\\
\kpijjj &= \kpi + x_i (\qonp + \qtwp + \qthp) - \delta_{ij}\qonp - \delta_{ij'}\qtwp - \delta_{ij''}\qthp\,.
\end{split}
\ee
We note that $G_{3\Lambda'\Lambda}$
is symmetric under the exchange of any two of its transverse momentum arguments.
Also, when $\Lambda=\Lambda'$ ($\Lambda\ne \Lambda'$), it is even (odd) under
a simultaneous sign flip of all three $\qpi$.

In this LCwf model the helicity transfer between the two proton states
is made possible by the quark orbital angular momentum originating
from the Melosh rotation $\sim\boldsymbol{\sigma}_\perp\times\boldsymbol{k}_\perp$, where $\boldsymbol{\sigma}_\perp = (\sigma^1,\sigma^2)$ are the transverse Pauli matrices, cf.\ ref.~\cite{Pasquini:2010af}. This is a purely kinematical effect, connecting the quark spin to the orbital motion in the relativistic formulation and unrelated to the conventional spin-orbit coupling $L^z S^z$ which is of dynamical origin.  The proton wave
functions $\Psi_\Lambda(x_i,\kpi,\lambda_i)$ contain quark helicity
contributions that go beyond the static quark-model spin structure.
For example, a helicity $+1/2$ proton may correspond to quark
helicities $(+1/2,-1/2,-1/2)$, since the mismatch is compensated by
quark orbital angular momentum -- the resulting proton wave
function is a $P$-wave (see the fourth line in \eqref{eq:spinwf}). As
the eikonal operator in \eqref{eq:rho3} does not flip the quark
helicities, the proton helicity flip matrix element
$G_{3-+}(\qonp,\qtwp,\qthp)$ relies on wave function overlap
that is diagonal in quark helicities, and the outgoing proton with
helicity $-1/2$ is in an $S$-wave. In the full expression, the
overlaps contain not only $S$-wave and $P$-wave interferences, but
also $P$-wave and $D$-wave interferences. We note the distinction between the off-diagonal matrix element $G_{3 -+}$ sourcing the gluon Sivers function and the matrix element for the orbital angular momentum distribution (see e.~g.~\cite{Lorce:2011ni}) which is, by contrast, diagonal in the proton helicity and, in the LCwf model, diagonal in the total parton orbital angular momentum \cite{Lorce:2011ni}.
%}

For forward scattering the first and second terms in eq.~(\ref{eq:g3}), where
the three $t$-channel gluons attach to one or two quarks in the proton, respectively, vanish when $\Lambda' \ne \Lambda$. It is impossible to
transfer orbital angular momentum to one or two quarks when the total
momentum transfer vanishes. Hence, it is required to consider a system of
at least three target partons.
Mathematically this comes about in eq.~(\ref{eq:g3}) 
due to the odd parity of the helicity wave function overlap
$\Psi^*_{\Lambda'}(x_i, \kpi,\lambda_i)\, \Psi_{\Lambda}(x_i, \kpi,\lambda_i)$
under a simultaneous sign flip of all quark transverse momenta $\kpi$;
see appendix~\ref{sec:LCwf-qqq} for explicit expressions for the
LC helicity wave functions.
For this same reason, the two-gluon exchange matrix element with helicity
flip requires non-zero momentum transfer $\Delta_\perp$~\cite{Benic:2025ral}.
\\

As described thus far, eq.~(\ref{eq:dipole}) now pertains to scattering of a
dipole from the soft gluon field sourced by the valence quarks
eq.~(\ref{eq:|P>}).  To improve this result, one may consider the emission of
a gluon with moderate $x$ from one of the quarks which adds to
the fluctuating color charge sources in the proton. Such
computations have been performed
elsewhere~\cite{Dumitru:2020gla,Dumitru:2021tqp}, albeit for
simplified helicity wave functions which did not allow non-zero
helicity flip matrix elements of the eikonal dipole.  For
moderate $x\simeq 0.1$ and dipole sizes $r_\perp \simeq 0.1-0.5$~fm,
and $\alpha_s \lesssim 0.3$, the corrections have been found to
amount to less than a factor of 2. Fixed order perturbative corrections
to the helicity flip matrix elements are presently not known
but are expected to be of similar magnitude.
Finally, we mention that the emission of soft gluons with momentum
fractions much less than those of the valence partons shall be
re-summed into the RG-evolved source via the BFKL
equation in the following subsection.

\subsection{BFKL evolution of the helicity flip Odderon}
\label{sec:BFKLevol}

With increasing energy or decreasing $x$ the phase space for gluon emissions
opens up. It becomes necessary then to resum pairwise interactions, i.e.\ gluon
exchanges among the three gluons exchanged by the dipole and the proton.
Such resummation is achieved by the small-$x$ QCD evolution
equation. The non-linear evolution equation for the Odderon(s) has been derived in refs.~\cite{Kovchegov:2003dm,Hatta:2005as}. Here, however, we only require
the linear BFKL evolution~\cite{Lipatov:1976zz,Kuraev:1977fs,Balitsky:1978ic}, since at the initial $x_0$ both the
two- and three-gluon exchange amplitudes (for a perturbative dipole, $r_\perp
\ll \Lambda_\mathrm{QCD}^{-1}$) are small: 
${\cal P}_{\Lambda'\Lambda}(\rp)$, ${\cal O}_{\Lambda'\Lambda}(\rp) \ll 1$. Non-linear effects in the evolution equation act to decrease the Odderon amplitude \cite{Yao:2018vcg,Motyka:2005ep,Benic:2024pqe}, however this only plays a role when scattering off the proton approaches
the unitarity limit, ${\cal P}(\rp) \to 1$. 
(The evolution with energy of the $C$-even scattering amplitude
of the LC quark model has been studied in ref.~\cite{Dumitru:2023sjd}.
For a comparison of linear vs.\ non-linear evolution of the spin
independent Odderon see ref.~\cite{Benic:2024pqe}.)
Therefore, we expect that the quantitative impact on Odderon related observables will be small for perturbative dipoles, at the energies available at the future EIC.

In the leading $\log 1/x$ approximation, in the framework of
Mueller's dipole approach~\cite{Mueller:1993rr} the evolution equation formulated in position space reads
\begin{equation}
    \frac{\partial}{\partial Y} {\cal O}_{\Lambda'\Lambda}(\xp,\yp) =
    \frac{\bar\alpha_s}{2\pi}\int_{\zp} \, 
    \frac{(\xp-\yp)^2}{(\xp-\zp)^2\, (\zp-\yp)^2}\,
    \left[ {\cal O}_{\Lambda'\Lambda}(\xp,\zp) + {\cal O}_{\Lambda'\Lambda}(\zp,\yp) - {\cal O}_{\Lambda'\Lambda}(\xp,\yp)
    \right]~.
\end{equation}
Here, $\bar\alpha_s = \alpha_s N_c/\pi$ and $Y=\log x_0/x$ where $x_0 \sim 0.1$
corresponds to the initial condition.
For dipole scattering with Odderon exchange, the above is equivalent
to the Bartels-Kwiecinski-Praszalowicz (BKP) linear evolution equation~\cite{Bartels:1980pe,Kwiecinski:1980wb}, see the discussion in
ref.~\cite{Kovchegov:2003dm}.

Assuming translational invariance, ${\cal O}_{\Lambda'\Lambda}(\xp,\yp) \to
{\cal O}_{\Lambda'\Lambda}(\xp-\yp)$,
and integrating over the impact
parameter $\bp = (\xp+\yp)/2$ (i.~e.~taking $\delp \to 0$), we arrive at the evolution equation for the
forward matrix element,
\begin{equation}
    \frac{\partial}{\partial Y} {\cal O}_{\Lambda'\Lambda}(\rp) =
    \frac{\bar\alpha_s}{2\pi}\int_{\ssp} \, 
    \frac{\rp^2}{\left(\frac{\rp}{2}-\ssp\right)^2\, \left(\frac{\rp}{2}+\ssp\right)^2}\,
    \left[ {\cal O}_{\Lambda'\Lambda}\left(\frac{\rp}{2}-\ssp\right) + 
    {\cal O}_{\Lambda'\Lambda}\left(\frac{\rp}{2}+\ssp\right) - 
    {\cal O}_{\Lambda'\Lambda}(\rp)
    \right]~.
    \label{BFKL_imindep}
\end{equation}
We note that aside from the trivial fixed point ${\cal O}_{\Lambda'\Lambda}(\rp) =0$, the equation also possesses a non-trivial fixed point for an Odderon
that is linear in the dipole size: ${\cal O}_{\Lambda'\Lambda}(\rp) \sim
\Sp \times \rp$. To our knowledge this fixed point of the helicity flip Odderon
has not been pointed out before.
Deep in the perturbative regime of very small dipoles, however, the
initial condition
will be of cubic order in $r_\perp$, as already alluded to above.
The BFKL resummation is expected to generate an anomalous dimension 
$\gamma(Y)<1$ which we determine numerically in sec.~\ref{sec:Results-Discussion} below.

Equation (\ref{BFKL_imindep}) can be simplified using translational and rotational invariance: we fix $\rp = (r_\perp \cos\phi_r,r_\perp\sin\phi_r) = (r_\perp,0)$ in the polar coordinates. The angular dependence of ${\calO}_{\Lambda'\Lambda}(\rp)$ is given by \eqref{eq:Odecompose} \footnote{Similar formulas hold for $\mathcal{O}_{\Lambda'\Lambda}(\kp)$.}:
\begin{equation}
\calO_{-+}(\rp) = \rmi \rme^{\rmi\phi_r}\calO_S(r_\perp)\,.
\end{equation}
where the radial part $\calO_S(r_\perp)$ is real and satisfies the equation:
\begin{equation}
\frac{\partial}{\partial Y} \calO_S(Y,r_\perp) = \frac{\bar{\alpha}_s}{2\pi} \int_{\zp}\frac{\rp^2}{ (\rp - \zp)^2 \zp^2} \Biggl[ \frac{r_\perp - z_\perp\cos\phi_z}{|\rp - \zp|} \calO_S(Y, |\rp - \zp|) +  \cos\phi_{z} \calO_S(Y,z_\perp) - \calO_S(Y, r_\perp) \Biggl]\,.
\label{odd_ev_equation_Im}
\end{equation}
Here we explicitly indicate that $\calO_S$ depends on the rapidity $Y$;
the initial condition at $Y=0$ is given by the perturbative three gluon exchange of eq.~(\ref{eq:oddmixed}) with $\Lambda'
= - \Lambda$ and $\Delta_\perp=0$, using the three gluon to proton coupling
$G_{3\Lambda'\Lambda}$ obtained via eq.~(\ref{eq:g3})
from the non-perturbative model LCwf. 
In the numerical implementation, we shall use $\alpha_s=0.25$.
At the current order
the coupling enters the Sivers function only as an overall
factor, $x f_{1T}^\perp (x,k_\perp) \sim \alpha_s^2$, and as a scale
factor in the evolution ``time" $t\sim \alpha_s Y$.

%--------------------------------------------------------
\section{Numerical results}
\label{sec:Results-Discussion}

In this section we present and discuss our numerical results. We shall first
address the helicity flip Odderon and the associated gluon Sivers function
obtained from the perturbative computation of three gluon exchange.
Then, we illustrate the properties of that Sivers function (such as its magnitude,
sign definiteness etc.) by the example of a specific physical process.
Lastly, we show the effect of BFKL resummation of contributions
$\sim (\alpha_s \log x_0/x)^n$.
The figures shown in this section were obtained with the
exponential quark-level LCwf, see appendix~\ref{sec:LCwf-qqq}.
We have verified that the alternative power-law wave function of
app.~\ref{sec:LCwf-qqq} produced very similar results.
\\

\begin{figure}[ht]
    \centering
    \begin{subfigure}[t]{0.31\linewidth}
        \centering
        \includegraphics[width=\linewidth]{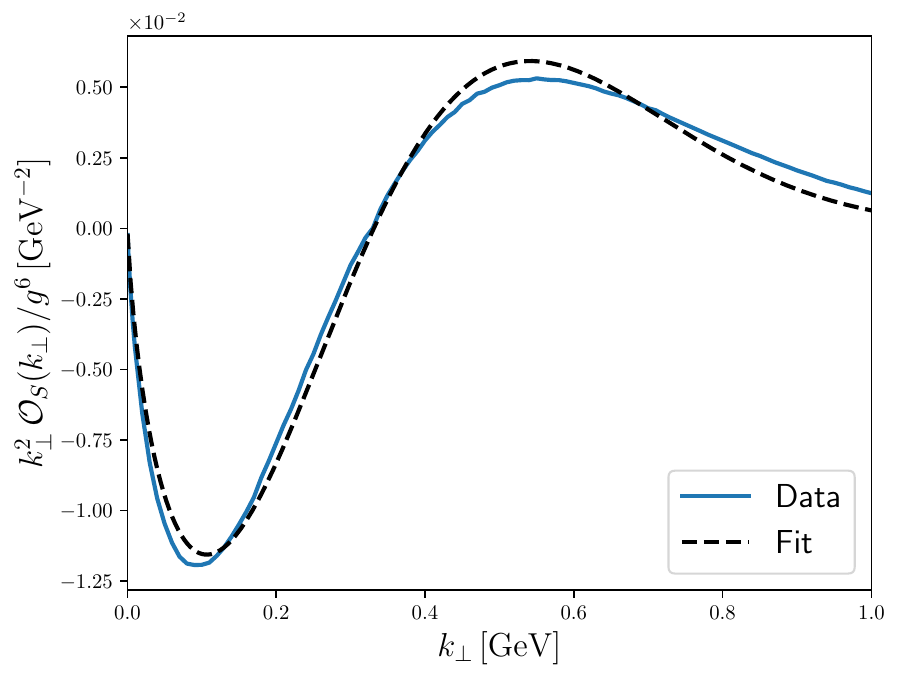}
        \caption{}
        \label{fig:odderon_dist_vegas}
    \end{subfigure}
    \hfill
    \begin{subfigure}[t]{0.34\linewidth}
        \centering
        \includegraphics[width=\linewidth]{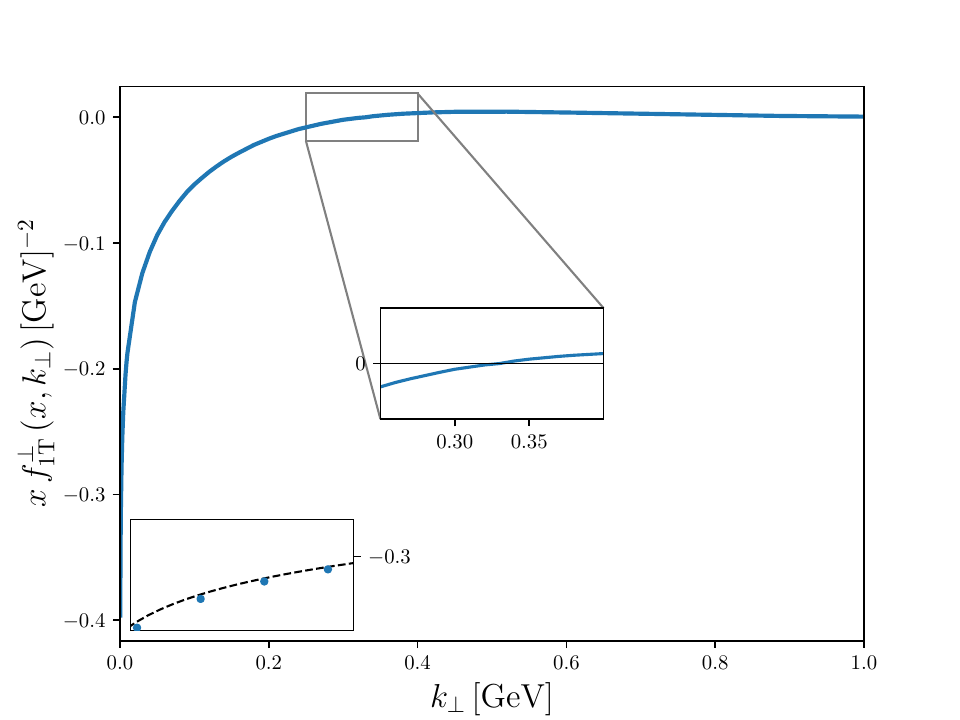}
        \caption{}
        \label{fig:gluon_sivers_vegas}
    \end{subfigure}
    \hfill
    \begin{subfigure}[t]{0.31\linewidth}
        \centering
        \includegraphics[width=\linewidth]{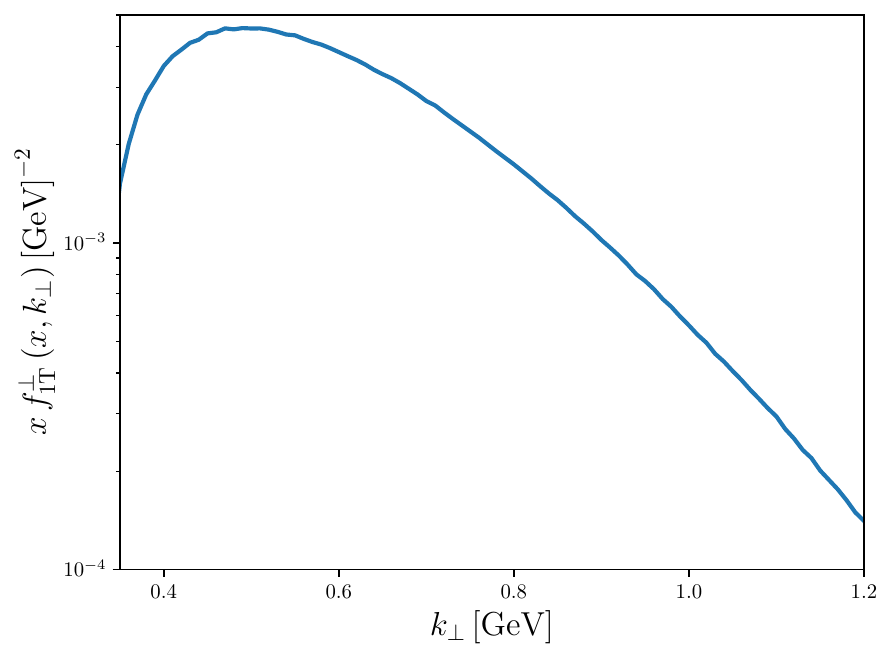}
        \caption{}
        \label{fig:gluon_sivers_vegas_pos}
    \end{subfigure}
    \caption{(a) The forward Odderon $k_\perp^2\calO_S(k_\perp)/g^6$ in transverse momentum space.  The dashed line shows the best fit via eq.~\eqref{eq:k2odderon_log_fit}. (b) and (c) Gluon Sivers function  $x f_{1T}^\perp (x,k_\perp)$
    obtained from the helicity flip, forward Odderon for $\alpha_s =0.25$. The dashed black line in the lower inset
    %\sout{corresponds to the logarithmic fit of eq.~\eqref{eq:sivers_log_fit}}
    emphasizes the log-type divergence at low-$k_\perp$, see also \eqref{eq:k2odderon_log_fit}
    %}
    .}
    \label{fig:odderon_dist_gluon_sivers}
\end{figure}
Fig.\ref{fig:odderon_dist_vegas} shows the transverse momentum
dependence of the forward, helicity flip Odderon at $x_0$.
We may guess the following functional behavior of 
the helicity flip Odderon:
%\begin{equation}
%\label{eq:k2odderon_log_fit}
%    k_\perp^2\, \frac{\left|\mathcal{O}_{-+}(\kp)\right|}{g^6} = \mathcal{N} \left(\frac{k_\perp}{Q_b}\right)^a \log\left(\frac{k_\perp}{Q_b}\right) e^{-k_\perp^2/(4Q_c^2)} \,.
%\end{equation}
\begin{equation}
\label{eq:k2odderon_log_fit}
    k_\perp^2\, \frac{\left|\mathcal{O}_{-+}(\kp)\right|}{g^6} = \mathcal{N} \frac{k_\perp}{Q_b} \log\left(\frac{k_\perp}{Q_b}\right) e^{-k_\perp^2/(4Q_c^2)} \,.
\end{equation}
The best fit parameters for the exponential wave function in appendix \ref{sec:LCwf-qqq} are
%In particular, for the result in Fig.\ref{fig:odderon_dist_vegas} we find 
% $\mathcal{N}=0.029\text{ GeV}^{-2},\ a=0.901,\ Q_b=0.333\text{ GeV},\ Q_c=0.232\text{ GeV}$.
$\mathcal{N}=0.036\text{ GeV}^{-2},\ Q_b=0.331\text{ GeV},\ Q_c=0.220\text{ GeV}$.
Note that the fit is not meant to
describe the analytic form
% \sout{of $k_\perp^2\mathcal{O}(k_\perp)$ for $k_\perp \to 0$;also, it should not be extended}
beyond $k_\perp \simeq 1$~GeV.

Fig.~\ref{fig:gluon_sivers_vegas} shows the Sivers function
in the regime of low $k_\perp$. The inset depicts its logarithmic divergence
for very low transverse momenta which is evident from \eqref{eq:k2odderon_log_fit}.
Such logarithmic behavior at low $k_\perp$ is not
manifest in the Sivers function from the eikonal, ``cubic Casimir extended"
MV model~\cite{Zhou:2013gsa,Yao:2018vcg}. 
It arises from eq.~(\ref{eq:O(kT)}),
in the $\Delta_\perp \to 0$ limit, if the 
three-gluon exchange is dominated by the exchange of two fairly hard gluons
with nearly anti-collinear transverse momenta, and a softer third gluon;
and if in that regime the helicity flip
three gluon coupling to the proton is $G_{3-+}(\kp,\qtwp,-\qtwp) \sim
k_L \, q_{2R} \, q_{2L} = k_L \, \qtwp^2$\footnote{$q_{R/L} \equiv q_x \pm \rmi q_y$.}. The integration over $\qtwp$ will then result in a logarithmic
behavior of the Sivers function.
Intuitively, such behavior of $G_{3-+}$ which corresponds
to the exchange of vector gauge bosons, appears natural. 
It is similar to the way in which the electromagnetic form
factor $F_{+-}$ is proportional to the transverse momentum of the exchanged photon, see eq.~(\ref{eq:F12}). However, this point
may warrant more detailed investigation in the future.

Since that $x f_{1T}^\perp (x,k_\perp)$, in 
the non-perturbative regime of low $k_\perp$,
may not be suppressed $\sim k_\perp^2$ but may actually diverge logarithmically, therefore caution is advised with observables such as single spin asymmetries which
integrate $x f_{1T}^\perp (x,k_\perp)$ over $k_\perp$. On the other hand,
the collinear function $xf_{1T}^{\perp (1)}(x,\mu^2)$ is safe from
logarithmic low-$k_\perp$ contributions, see
eq.~(\ref{eq:f-Siv-kt2-moment}).

The behavior for higher $k_\perp$ is shown in fig.~\ref{fig:gluon_sivers_vegas_pos}, on a logarithmic scale. The Sivers function peaks
at about $k_\perp \simeq 0.5$~GeV where its magnitude is
$(4-5)\times 10^{-3}$~GeV$^{-2}$. In the neighborhood of $k_\perp = 1$~GeV
it falls off from a magnitude of about $10^{-3}$~GeV$^{-2}$
at $k_\perp = 0.9$~GeV to about $10^{-4}$~GeV$^{-2}$
at $k_\perp = 1.2$~GeV. Beyond that, the function is expected to exhibit a
perturbative power-law tail $\sim 1/k_\perp^{4}$ which we
mention below in conjunction with BFKL-resummed solutions.
\\

\begin{figure}
    \centering
    \includegraphics[width=0.5\linewidth]{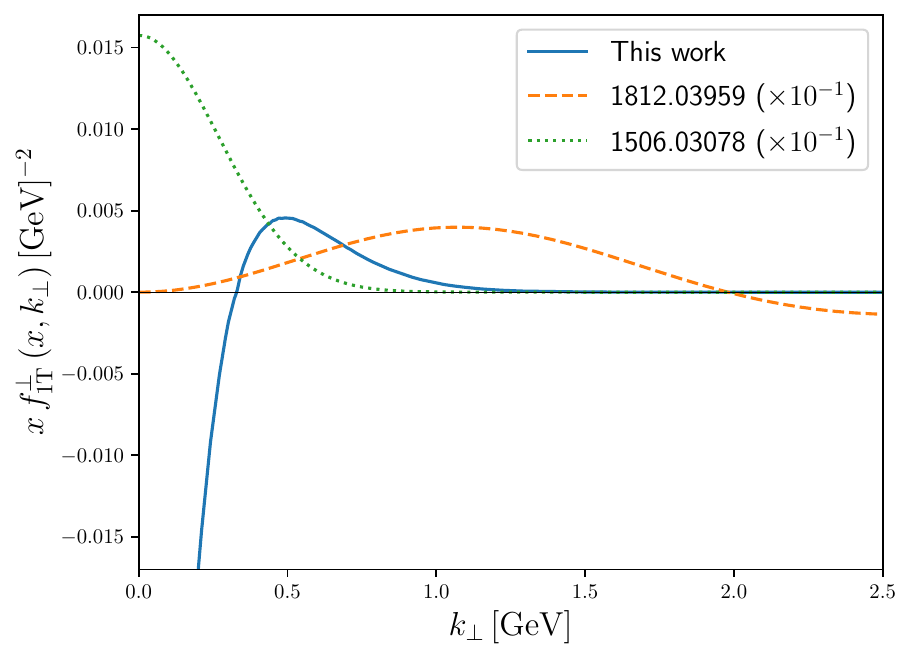}
\caption{Comparison of the eikonal gluon Sivers function $xf_{1T}^\perp(x,k_\perp)$ to the fit and model of Refs.~\cite{DAlesio:2015fwo} (green dotted) and \cite{Yao:2018vcg} (orange dashed), respectively, with adjusted scales for easier comparison.}
    \label{fig:sivers_comparison}
\end{figure}
In Fig.~\ref{fig:sivers_comparison} we compare our model prediction for the eikonal gluon Sivers function $xf_{1T}^\perp(x,k_\perp)$ to other models from the literature. 
In particular, Ref.~\cite{DAlesio:2015fwo} performed a fit to transverse single spin asymmetry measurements in $pp\to\pi^0X$ at RHIC \cite{PHENIX:2013wle}. Their ``SIDIS2" fit involves an {\it ansatz} based on the generalized parton model approach whereby $f_{1T}^\perp(x,k_\perp)$ is parametrized in terms of the unpolarized collinear gluon distribution $f(x)$ times a Gaussian in $k_\perp$, and constrained by the Burkardt sum rule~\cite{Burkardt:2004ur}. 
This trivially satisfies the positivity bound ($f_{1T}^\perp(x,k_\perp)\leq 2 f(x)$), albeit at the cost of lacking a node in $k_\perp$. 
On the other hand, the low-$x$ dipole approach of Ref.~\cite{Yao:2018vcg} leaves the sign and overall magnitude arbitrary but explicitly enforces a node via the initial condition $\mathcal{O}_S(r_\perp)\sim r_\perp^3 e^{-r_\perp^2 Q_{s0}^2}$ for the dipole Odderon, where $Q_{s0}=0.7$~GeV was used for the saturation scale. 
Since the node structure in $k_\perp$ is unchanged by BFKL evolution,
we restrict our comparison in Fig.~\ref{fig:sivers_comparison} to the initial conditions at $Y=0$.
Lastly, we note that around $k_\perp \simeq 0.4$~GeV the Sivers function obtained
here is suppressed by about one order of magnitude as compared to the
model by D'Alesio {\em et al.}~\cite{DAlesio:2015fwo}, albeit this discrepancy diminishes with increasing
$k_\perp$.
\\

We scrutinize the phenomenological implications of
the Sivers function we obtained by considering a specific
physical process, i.e.\ the exclusive forward
production of $\chi_{c1}$ axial vector mesons on unpolarized protons~\cite{Benic:2024fbf}.
(Other observables for the spin dependent Odderon have been mentioned in
the Introduction, and shall be considered in the future.)
In the non-relativistic limit this process probes the tri-gluon PDF of the proton \eqref{eq:f-Siv-kt2-moment}.
However, it occurs both due to spin-dependent Odderon as well as due to photon
exchange, a Primakoff-like process. For an axial-vector state such as the
$\chi_{c1}$ the latter cross section does not exhibit the usual
Coulomb singularity at $|t| \to 0$ thanks to the Landau-Yang theorem \cite{Benic:2024pqe}. The finite ratio
of three-gluon versus photon exchange cross sections in the non-relativistic
heavy-quark and collinear limits is~\cite{Benic:2024fbf}
\be
r(x,\mu^2) = \frac{4\pi^2}{q_c^2 N_c^2} \frac{\alpha_s^2}{\alpha^2}
\frac{M_p^2}{M_\chi^2} \, \left| xf_{1T}^{\perp(1)g}(x,\mu^2)
\right|^2~.
\ee
The scale $\mu$ should be in the range of $M_\chi/2 - M_\chi$, and $Q^2 =0$
has been assumed.
Our numerical evaluation of the function $xf_{1T}^{\perp(1)g}(x,\mu^2)$
results in a ratio to Primakoff of only $r\approx 10^{-6}$.
This very small value of $r$ at $x \simeq 0.1$ is due to substantial
cancellations in the $k^2_\perp$ moment~(\ref{eq:f-Siv-kt2-moment}) of the 
Sivers function which satisfies the sum rule~(\ref{eq:f-Siv-sum-rule}).
Based on the ``SIDIS2" and ``5\% positivity" models of refs.~\cite{DAlesio:2015fwo}
and~\cite{Zheng:2018ssm}, respectively,
the $r$-ratio estimate of ref.~\cite{Benic:2024fbf} at $x \simeq 0.1$
was 3 to 4 orders of magnitude higher.
The small magnitude of the tri-gluon PDF obtained here implies that
forward exclusive photoproduction produces transversely polarized 
$\chi_{c1}$~\cite{Benic:2024fbf}.
\\ 

We now proceed to discuss the solution of
the BFKL equation.
The above initial condition exhibits long tails towards large dipoles which
complicate the numerical solution of eq.~(\ref{odd_ev_equation_Im}).
Therefore, we multiply the initial condition (\ref{eq:oddmixed}) by a Gaussian suppression factor:
\begin{equation}
\calO_S(Y=0,r_\perp) \to \rme^{-r_\perp^2 \mu^2} \calO_S(Y=0,r_\perp)\,. 
\end{equation}
This ``cutoff" suppresses contributions from large $r_\perp$ which correspond to
non-perturbative, large dipoles. We choose $\mu=0.1$~GeV, so that our $\calO_S(Y=0,r_\perp)$ is not altered much for $r_\perp<5$~GeV$^{-1}$, which is the region we shall focus on. 

\begin{figure}[htbp]
    \begin{subfigure}[t]{0.49\linewidth}
        \centering
        \includegraphics[width=\linewidth]{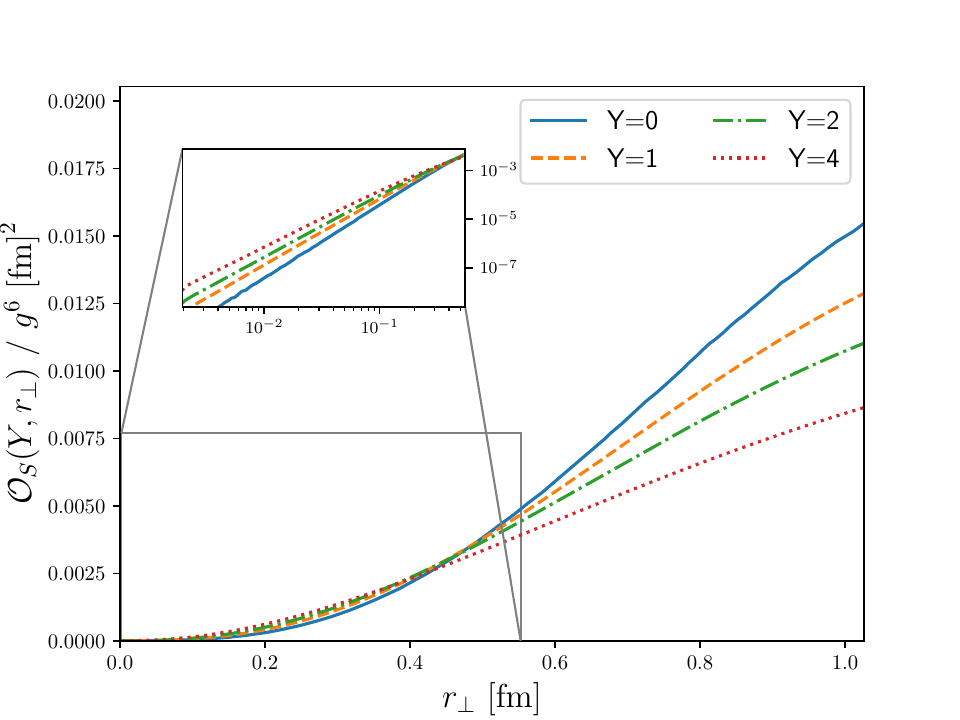}
        \caption{}
        \label{fig:evol_O_r}
    \end{subfigure}
    \begin{subfigure}[t]{0.49\linewidth}
        \centering
        \includegraphics[width=\linewidth]{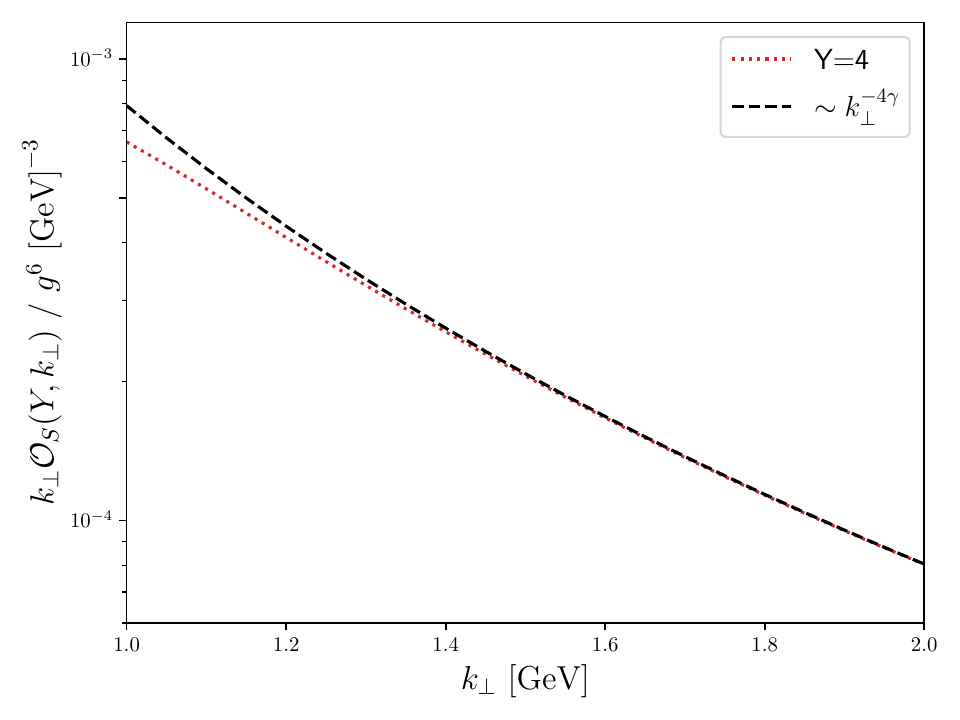}
        \caption{}
        \label{fig:evol_O_k}
    \end{subfigure}
\caption{(a) Radial part of the helicity flip Odderon amplitude at different rapidities, $Y=0, 1, 2, 4$. The inset is a double logarithmic version of the plot to better show the small-$r_\perp$ regime. (b): high-$k_\perp$ tail of the momentum-space Odderon amplitude $k_\perp \calO_S(k_\perp)$, which is proportional to $xf_{1T}^{\perp g}(x,k_\perp^2)$, at $Y=4$. 
The dashed line indicates the $\sim1/k_\perp^{3.3}$ power-law asymptotics.}
\label{fig:evol_O}
\end{figure}

In Fig.~\ref{fig:evol_O_r}, we show (the radial part of) 
the helicity flip Odderon amplitude for several values of rapidity $Y$: 0 (solid blue), 1 (dashed orange), 2 (dsh-dotted green) and 4 (dotted red). We observe that small-$x$ evolution increases
$\calO_S(Y,r_\perp)$ at small $r_\perp \lesssim 0.45$~fm, while it 
suppresses it for greater $r_\perp \gtrsim 0.45$~fm; this point corresponds
to the fixed point of linear $\calO_S(r_\perp)$ mentioned above. 
The suppression
at moderately large $r_\perp$ was already noted in ref.~\cite{Yao:2018vcg}. 
We find that the asymptotic behavior at small $r_\perp$, $\calO_S(Y,r_\perp) \sim r_\perp^{3\gamma(Y)}$, changes with evolution: the anomalous dimension $\gamma(Y)$ decreases with $Y$, 
starting from $\gamma(Y=0)=1$ for
perturbative three gluon exchange and reaching $\gamma(Y=4)\approx 0.8$.

Fig.~\ref{fig:evol_O_k} shows the high-$k_\perp$ power-law tail of the
spin dependent Odderon times $k_\perp$; this is proportional to the gluon
Sivers function $xf_{1T}^{\perp g}(x,k_\perp^2)$. At the initial rapidity
$Y=0$, to exhibit the power-law tail of the perturbative three gluon exchange, we require a small anomalous dimension
$1-\epsilon$ for $\calO_S(r_\perp)$; 
this then leads to $xf_{1T}^{\perp g}(x,k_\perp^2) \sim 1/k_\perp^4$.
Comparing to unpolarized gluon TMD with a perturbative $\sim 1/k_\perp^2$ tail, the stronger fall-off of the gluon Sivers function is due to its connection to the Odderon, which is a three gluon exchange. This is similar to diffractive TMDs which are described by four gluon exchanges, also leading to a stronger 
$\sim 1/k_\perp^4$ tail \cite{Iancu:2022lcw,Hauksson:2024bvv}.
%}

At $Y=4$ (for $\alpha_s=0.25$) the power-law tail of the solution of BFKL becomes $\sim 1/k_\perp^{4\gamma(Y)}$ with $\gamma(Y) \approx 0.8$. This value of the anomalous dimension agrees with the position space behavior of $\calO_S(Y,r_\perp)$ at small $r_\perp$.
For asymptotic rapidity, $\gamma(Y)$ approaches the standard LO BFKL
anomalous dimension, $\gamma=0.5$. Numerically, we find that this
requires $Y \gtrsim 16$ where non-linear unitarity effects are expected
to become important.
Hence, we find it potentially more relevant to quote
the value of $\gamma$ at $\alpha_s Y =1$ where BFKL evolution has made a
substantial yet pre-asymptotic effect on the high-$k_\perp$ tail.

%----------------------------------------------------------------------

%----------------------------------------------------------------------
\section{Summary and discussion}

To date the most commonly used parameterization for the gluon Sivers function
of the proton is the model by D'Alesio {\it et al.}~\cite{DAlesio:2015fwo}.
However, this model does not originate from the correspondence at small $x$ to the
helicity flip Odderon, and is not well suited as an initial condition for
BFKL evolution. Hence, it is not straightforward to determine from it
the BFKL anomalous dimension for the small-$x$ gluon Sivers function.

Our goal here was to develop an alternative based on the eikonal
approach, and on an effective quark model LCwf
for the proton. Such a model incorporates empirical knowledge of the proton
at moderate energy and transverse resolution scales, and can be tuned to
provide a satisfactory description of Dirac and Pauli form factors.
In particular, the model specifically involves sources (for the soft
$t$-channel gluons exchanged in high-energy scattering) in the fundamental
representations of both color-SU(3) and spin-SU(2).

The above light-cone quark model for the proton enables the explicit computation of
the matrix element of the eikonal dipole operator between proton states of
opposite helicity. While the $C$-even part of this matrix element requires
non-zero momentum transfer~\cite{Benic:2025ral},
the $C$-odd part is finite
even in the limit of forward scattering, where it defines the ``spin dependent
Odderon" and the gluon Sivers function~\cite{Boer:2015pni,Zhou:2013gsa,Yao:2018vcg,Boussarie:2019vmk},
a transverse momentum dependent parton distribution.
\\

Our main results for the small-$x$ dipole gluon Sivers function are as follows.
We confirm the statements in refs.~\cite{Zhou:2013gsa,Yao:2018vcg} that
$xf_{1T}^{\perp}(x,k_\perp)$, as a function of transverse momentum $k_\perp$,
is not sign definite. This leads to substantial cancellations in the first
$k_\perp^2$ moment of the Sivers function, $xf_{1T}^{\perp (1)}(x,\mu^2)$,
which is related to the twist three collinear function frequently denoted the
``tri-gluon PDF" $O(x,\mu^2)$; this is true
especially for relatively low scales $\mu$ of order $1-3$~GeV.
A related phenomenological consequence is that forward exclusive photoproduction
of $\chi_{c1}$ axial vector mesons is dominated by photon rather than
Odderon exchange~\cite{Benic:2024fbf}.

We provide the following parameterization of the eikonal gluon Sivers function
at $x = x_0 \sim 0.1$:
\begin{equation}
\label{eq:kTxf(kT)_fit}
    k_\perp\, xf_{1T}^{\perp}(x,k_\perp) =
    % \frac{6 M_p}{(2\pi)^4\, \alpha_s}\,
    % include factor of g^6
    \frac{24 M_p}{\pi}\, \alpha_s^2\,
    \mathcal{N} \frac{k_\perp}{Q_b} \log\left(\frac{k_\perp}{Q_b}\right) e^{-k_\perp^2/(4Q_c^2)} \,.
\end{equation}
The fit parameters are $\mathcal{N}=0.036\text{ GeV}^{-2},\ Q_b=0.331\text{ GeV},\ Q_c=0.220\text{ GeV}$.\footnote{Allowing the exponent of $(k_\perp/Q_b)\to(k_\perp/Q_b)^a$ to vary as well, we obtain a statistically more significant fit with $\mathcal{N}=0.029\text{ GeV}^{-2},\ a=0.901,\ Q_b=0.333\text{ GeV},\ Q_c=0.232\text{ GeV}$, albeit at the cost of an additional parameter.} This fit does not apply beyond $k_\perp \sim 1$~GeV.
%or for $k_\perp \to 0$.

The gluon Sivers function in this model exhibits a peak (at $k_\perp>Q_b$)
at a transverse momentum scale typical of the quarks in the proton. 
This is unlike the parameterization of $xf_{1T}^{\perp}(x,k_\perp)$ by D'Alesio 
{\it et al.}~\cite{DAlesio:2015fwo} which is a Gaussian, and has its maximum
at $k_\perp = 0$. The peak obtained in our model resembles the one found in
saturation models of the Odderon~\cite{Yao:2018vcg} albeit its physical origin
in our purely perturbative computation of three gluon exchange is rather 
different. The forward matrix element, in the proton, of the {\em eikonal} dipole operator 
for reversed helicity, involves orbital angular momentum (OAM) 
of the sources of the soft gluons:
despite the fact that the eikonal dipole operator conserves the
helicities of the individual quarks, a proton helicity flip is possible
due to non-vanishing overlap of quark configurations which differ by their orbital angular momentum.
The peak of $xf_{1T}^{\perp}(x,k_\perp)$ corresponds to the scale where the
transfer of OAM between initial and final proton state is most efficient.
Numerically, for the model LCwf used here the peak
occurs at about $k_\perp \simeq
0.5$~GeV where the magnitude of $xf_{1T}^{\perp}(x,k_\perp)$
is $(4-5)\times 10^{-3}$~GeV$^{-2}$, assuming
$\alpha_s=0.25$. 
While a comprehensive quantitative analysis of uncertainties
is beyond the scope of this paper, we do recall that ${\cal O}(\alpha_s)$
perturbative corrections to the non-perturbative three quark LCwf could
increase the magnitude of the gluon Sivers function at $x=x_0=0.1$ by up to
a factor of two. Rescaling the coupling changes $xf_{1T}^{\perp}(x,k_\perp)$
at that initial $x_0$ by a factor of $\alpha_s^2$. Lastly, variation of
the LC quark model wave function from exponential to power law, both
constrained by proton form factors, was found to modify
the Sivers function by factors of up to 1.5.

For small $k_\perp$ we obtain a logarithmic singularity
of the Sivers function rather than power-law decay~\cite{Yao:2018vcg}
or approach to a finite constant~\cite{DAlesio:2015fwo}. Hence, there could be
potentially significant contributions from small, non-perturbative transverse
momentum imbalances to charm single spin asymmetries. These would also be sensitive to
parton shower effects and to transverse momentum acquired in parton 
fragmentation~\cite{Zheng:2014vka, Boer:2022njw,Cassar:2025vdp,Caucal:2025zkl}. The resulting joint resummation framework is known to kinematically constrain small-$x$ radiations \cite{Beuf:2014uia,Caucal:2022ulg}, which would be important and interesting not only for quantitative phenomenology on single spin asymmetries for open charm at the EIC but also in view of the qualitative features of the gluon Sivers function found here. 
%(which are also sensitive to 
%parton shower effects and to transverse momentum acquired in parton 
%fragmentation~\cite{Zheng:2014vka,Cassar:2025vdp,Caucal:2025zkl}).
%We intend to investigate open charm production more quantitatively in the future.

The BFKL evolution of the spin dependent Odderon in this model generates
an anomalous dimension of its power-law tail. At high $k_\perp \gtrsim 1.5$~GeV
we find that $xf_{1T}^{\perp}(x,k_\perp) \sim k_\perp^{-4\gamma(Y)}$, with
$\gamma(Y)\simeq 0.8$ at $Y \equiv \log x_0/x = 1/\alpha_s$.
Also, given the above initial condition, we observe that
the magnitude of the perturbative small-$r$ tail of the
BFKL spin-dependent Odderon $\calO_S(Y,r_\perp)$ increases with $Y$ as $Y \to 1/\alpha_s$.
The non-forward, spin-{\em in}dependent dipole Odderon amplitude off the proton
also increases with decreasing $x$ if a gluon is added into the three quark
wave function~\cite{Dumitru:2022ooz}.
Asymptotically, the BFKL resummed Odderon amplitudes are expected to be 
(at most) independent of energy~\cite{Kovchegov:2003dm}.

\section*{Data and Code Availability}
To promote open and reproducible research, the data and computer code used to generate the numerical results and figures in this paper are publicly available via Zenodo \cite{zenodo15738460} and GitHub \cite{nucleon-lccqm}.

%-------------------------------------------------------------
\begin{acknowledgments}
  We thank Leszek Motyka for collaboration in the initial stages of this work and Edmond Iancu for comments on the manuscript.
  T.S.\ thanks  Piotr Korcyl for sharing his evolution code and Florian Cougoulic for discussions.
 A.D.\ acknowledges support by the DOE Office of Nuclear Physics through Grant 
 DE-SC0002307.
 F.H.\ is funded by the Austrian Science Fund (FWF) [10.55776/J4854].
 T.S.\ kindly acknowledges the support of the Polish National Science Center (NCN) Grant No.\,2021/43/D/ST2/03375.
\end{acknowledgments}

%*****************  appendices **************************
\appendix

\section{LC constituent quark model wave function}
\label{sec:LCwf-qqq}

In this appendix we present the detailed form of the three-quark LC wave function
$\Psi_\Lambda(x_i,\kpi,\lambda_i)$
used in our numerical evaluation of the Odderon.
We take the momentum space integrals in eq.~(\ref{eq:|P>}) as follows:
\be
[\rmd x_i] = 4\pi\delta(1-x_1-x_2-x_3) \prod_{i=1\cdots3} \frac{\rmd x_i}{4\pi\sqrt{x_i}}~,\qquad
[\rmd^2 \kpi] = (2\pi)^2 \delta(\konp+\ktwp+\kthp) 
\prod_{i=1\cdots3} \frac{\rmd^2 \kpi}{(2\pi)^2}~.
\ee
Next, we write $\Psi_\Lambda(x_i,\kpi,\lambda_i)$ as a product of a LC helicity
wave function and a spatial (momentum space) wave function,
\be
\Psi_\Lambda(x_i,\kpi,\lambda_i) = {\Phi}_\Lambda(x_i,\kpi,\lambda_i)\, {\psi}(x_i,\kpi) \,.
\label{eq:dec}
\ee
For the spatial wave function we use a simple model due to Schlumpf~\cite{Schlumpf:1992vq,Brodsky:1994fz}
whereby the suppression of the Fock space amplitude away from the minimal configuration is
governed by the invariant mass (squared) of the three-quark system,
$\mathcal{M}^2 = \sum_i ({\kpi^2+m^2})/{x_i}$. Schlumpf considered both a power-law
as well as an exponential wave function
\be
\psi_\mathrm{exp}(x_i,\kpi) = {\cal N_\mathrm{exp}}\,  e^{-\mathcal{M}^2/2\beta^2}~,\qquad
\psi_\mathrm{pwr}(x_i,\kpi) = {\cal N_\mathrm{pwr}}\, \left( 1+ \frac{\mathcal{M}^2}{\beta^2}\right)^{-p},~~~~~(p=3.5)~. 
\ee
The effective quark mass $m$ and the parameter $\beta$ have
been tuned in refs.~\cite{Schlumpf:1992vq,Brodsky:1994fz}
to reproduce the electromagnetic ``radius'' and the anomalous magnetic
moment of the proton.
In the subsequent appendix~\ref{sec:Dirac-Pauli-FF} we provide an update
of these parameters in view of more recent data on the electromagnetic form factors.
The normalization ${\cal N_{\textrm{exp}/\textrm{pwr}}}$ of
the spatial wave function follows from the requirement that
$\left<K,\Lambda' | P,\Lambda\right> = 16 \pi^3 P^+\delta(P^+ - K^+)\,
\delta^{(2)}(\Pp - \boldsymbol{K}_\perp) \delta_{\Lambda\Lambda'}$.
The results of sec.~\ref{sec:Results-Discussion} of the main text were obtained with the
exponential wave function. The power-law wave function did not produce
large differences in observables where the quark transverse momenta 
involved were limited to about $1-1.5$~GeV.

The helicity wave functions $\Phi_\Lambda(x_i,\kpi,\lambda_i)$
in~(\ref{eq:|P>}) for a proton with helicity $\Lambda/2$ are obtained
through a Melosh transformation of rest frame Pauli spinors to the
LF. Their explicit expressions for $\Lambda=+1$
are given in ref.~\cite{Pasquini:2008ax}, which we have confirmed.
For completeness, we also list them here:
\be
\begin{split}
 {\Phi}_\uparrow(\uparrow,\uparrow,\downarrow) & = \frac{1}{\sqrt{N_1 N_2 N_3}}\frac{1}{\sqrt{6}}\left(2 a_1 a_2 a_3 + a_1 k_{2L} k_{3R} + k_{1L} a_2 k_{3R}\right)\,,\\
 {\Phi}_\uparrow(\uparrow,\downarrow,\uparrow) & = \frac{1}{\sqrt{N_1 N_2 N_3}}\frac{1}{\sqrt{6}}\left(-a_1 a_2 a_3 + k_{1L} k_{2R} a_3 - 2 a_1 k_{2R} k_{3L}\right)\,,\\
 {\Phi}_\uparrow(\downarrow,\uparrow,\uparrow) & = \frac{1}{\sqrt{N_1 N_2 N_3}}\frac{1}{\sqrt{6}}\left(-a_1 a_2 a_3 + k_{1R} k_{2L} a_3 - 2 k_{1R} a_2 k_{3L}\right)\,,\\
 {\Phi}_\uparrow(\uparrow,\downarrow,\downarrow) & = \frac{1}{\sqrt{N_1 N_2 N_3}}\frac{1}{\sqrt{6}}\left(a_1 a_2 k_{3R} - 2 a_1 k_{2R} a_3 - k_{1L}k_{2R} k_{3R}\right)\,,\\
 {\Phi}_\uparrow(\downarrow,\uparrow,\downarrow) & = \frac{1}{\sqrt{N_1 N_2 N_3}}\frac{1}{\sqrt{6}}\left(a_1 a_2 k_{3R} - 2 k_{1R} a_2 a_3 - k_{1R}k_{2L} k_{3R}\right)\,,\\
 {\Phi}_\uparrow(\downarrow,\downarrow,\uparrow) & = \frac{1}{\sqrt{N_1 N_2 N_3}}\frac{1}{\sqrt{6}}\left(a_1 k_{2R} a_3 + k_{1R} a_2 a_3 + 2k_{1R} k_{2 R} k_{3L}\right)\,,\\
 {\Phi}_\uparrow(\uparrow,\uparrow,\uparrow) & = \frac{1}{\sqrt{N_1 N_2 N_3}}\frac{1}{\sqrt{6}}\left(- a_1 k_{2L} a_3 - k_{1L} a_2 a_3 + 2a_1 a_2 k_{3L}\right)\,,\\
 {\Phi}_\uparrow(\downarrow,\downarrow,\downarrow) & = \frac{1}{\sqrt{N_1 N_2 N_3}}\frac{1}{\sqrt{6}}\left(-a_1 k_{2R} k_{3R} - k_{1R} a_2 k_{3R} + 2 k_{1R} k_{2R} a_3\right)\,,\\
\end{split}
\label{eq:spinwf}
\ee
with $a_i = m + x_i \mathcal{M}$ and $N_i \equiv \kpi^2 + a_i^2$ and $(\uparrow,\downarrow)\equiv(+,-)$ in our notation.
These helicity wave functions pertain to the $|uud\rangle$ state; there is
no need here to consider quark flavor permutations since all the matrix
elements we are interested in are ``flavor blind".

The helicity wave functions for
$\Lambda=-1$ are obtained by a sign flip of $\Phi$, and the exchange
$k_{iL} \leftrightarrow - k_{iR}$, where $k_{R/L}=k_\perp^1 \pm \rmi k_\perp^2 = k_\perp e^{\pm \rmi \phi_k}$, and $\lambda_i \to - \lambda_i$.

\section{Dirac and Pauli electromagnetic form factors}
\label{sec:Dirac-Pauli-FF}

To exhibit the connection of the form factors to Generalized Parton Distributions (GPDs)
we start from their usual definitions
\be
P^+\int \frac{\rmd z^-}{2\pi} \rme^{i x P^+ z^-} \langle P' \Lambda' |\bar{\psi}(-z/2)\gamma^+ \psi(z/2)| P \Lambda\rangle = \bar{u} (P',\Lambda') \left[\gamma^+ H_q(x,\xi,\delp) + \frac{\rmi \sigma^{+\nu}\Delta_\nu}{2 M_p} E_q(x,\xi,\delp)\right] u(P,\Lambda)\,,
\ee
where $\Delta = P' - P$. We integrate this over $x$ and take the eikonal limit of
vanishing skewness, $\xi\to0$ or $P^{\prime +}\to P^+$:
\be
\langle P' \Lambda' |\bar{\psi}(0)\gamma^+ \psi(0)| P\Lambda\rangle = \bar{u} (P',\Lambda') \left[\gamma^+ F_1^q(\delp) + \frac{\rmi \sigma^{+\nu}\Delta_\nu}{2 M_p} F_2^q(\delp)\right] u(P,\Lambda)\,.
\label{eq:ffs}
\ee
where
\be
F_1^q = \int \rmd x H_q \,, \qquad F_2^q = \int \rmd x E_q\,,
\ee
are form-factors per flavor.

Eq.~\eqref{eq:ffs} involves a local quark current. Its Fourier transform at $q^+ \to 0$ defines
the eikonal LC charge operator
\be
\sum_f e_f\int \rmd z^- \int \rmd^2 \zp \rme^{\rmi q^+ z^- - \rmi \qp \cdot \zp} \bar{\psi}_f(z)\gamma^+ \psi_f(z) = \int \rmd z^- \rme^{i q^+ z^-} J^+(z^-,\qp) \approx \int \rmd z^- J^+(z^-,\qp) \equiv \rho(\qp)\,.
\ee
This charge operator can be written 
in terms of quark creation and annihilation operators~\cite{Dumitru:2018vpr}
in order to evaluate explicitly the matrix element on the l.h.s.\ of~\eqref{eq:ffs}.
This leads to
\be
\langle P' \Lambda' |\rho(\qp)| P \Lambda \rangle = 2 P^+ (2\pi)\delta(P'^+ - P^+) (2\pi)^2 \delta^{(2)}(\qp + \delp) F_{\Lambda'\Lambda}(\delp)\,,
\label{eq:emme}
\ee
where
\be
\begin{split}
F_{\Lambda'\Lambda}(\delp) & = \sum_{\lambda_i} \int_{x_i} 4\pi \delta(1 - x_1 - x_2 - x_3) \int_{\kpi} (2\pi)^2 \delta(\konp + \ktwp + \kthp)\\
&\times \Big[e_d\Psi_{\Lambda'}^{*}(x_i,\kpi - x_i \delp + \delta_{i3}\delp,\lambda_i)\Psi_{\Lambda}(x_i,\kpi,\lambda_i)\\
& + e_u\Psi_{\Lambda'}^{*}(x_i,\kpi - x_i \delp + \delta_{i2}\delp,\lambda_i)\Psi_{\Lambda}(x_i,\kpi,\lambda_i)\\
& + e_u\Psi_{\Lambda'}^{*}(x_i,\kpi - x_i \delp + \delta_{i1}\delp,\lambda_i)\Psi_{\Lambda}(x_i,\kpi,\lambda_i)\Big]\,.
\end{split}
\ee
corresponds to a kind of helicity form factor.
$F_{\Lambda'\Lambda}$ is related to $F_1$ and $F_2$ via
\be
F_{\Lambda'\Lambda}(\delp) = F_1(\delp)\delta_{\Lambda'\Lambda} + \frac{\Delta_\perp}{2M_p}F_2(\delp) \Lambda\rme^{\rmi \Lambda\phi_\Delta} \delta_{\Lambda,-\Lambda'}\,, 
\ee
or
\be
\begin{split}
% & F_1(\delp) = F_{\uparrow \uparrow}(\delp) = F_{\downarrow \downarrow}(\delp)\,,\\
% & \frac{1}{2M_p} F_2(\delp) \Delta_R = F_{\downarrow \uparrow}(\delp)\,,~~~~
% -\frac{1}{2M_p} F_2(\delp) \Delta_L = F_{\uparrow\downarrow }(\delp)\,.
& F_1(\delp) = F_{++}(\delp) = F_{--}(\delp)\,,\\
& \frac{1}{2M_p} F_2(\delp) \Delta_R = F_{-+}(\delp)\,,~~~~
-\frac{1}{2M_p} F_2(\delp) \Delta_L = F_{+-}(\delp)\,.
\end{split}
\label{eq:F12}
\ee
Importantly, the Pauli form factor of the {\em eikonal} current does not vanish thanks to the
fact that non-zero overlaps of the type $\Psi_{-\Lambda}^{*}(x_i,\kpi^\prime,\lambda_i)\, \Psi_{\Lambda}(x_i,\kpi,\lambda_i)$ exist. The helicity of the proton may flip due to the transfer of
orbital angular momentum to its constituents.
\\

\begin{figure}[htb]
  \begin{center}
  \includegraphics[width=.8\linewidth]{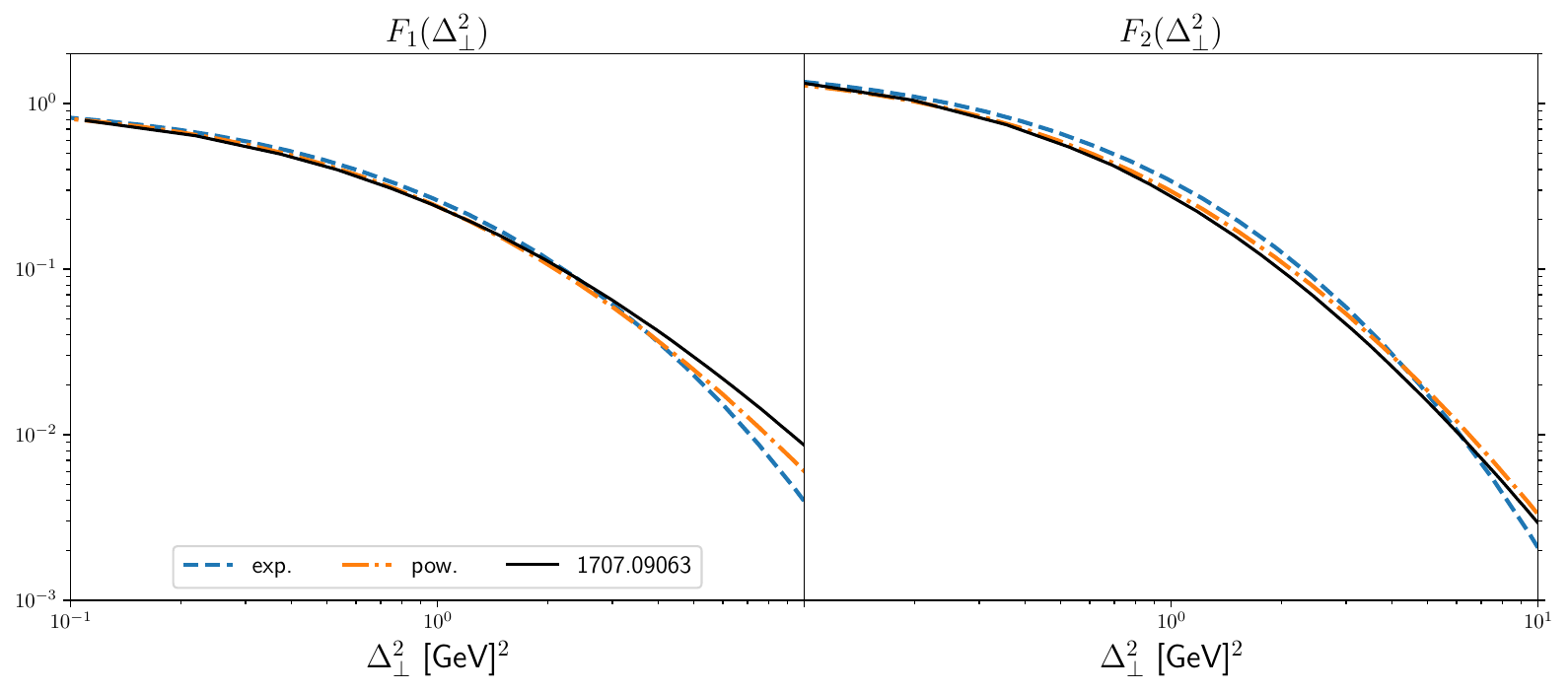}
  \end{center}
  \caption{The Dirac (left) and Pauli (right) electromagnetic form factors
    obtained from the three-quark LCwf compared to a modern fit (solid
    black line) to experimental data~\cite{Ye:2017gyb}.}
  \label{fig:Dirac-Pauli-FF}
\end{figure}
%Fig.~\ref{fig:Dirac-Pauli-FF} shows our best fit to the experimental parametrization
%from ref. \cite{Ye:2017gyb}, achieved by variation of the two parameters $\beta$ and $m$ in
%the three-quark LCwf. 
Fig.~\ref{fig:Dirac-Pauli-FF} shows our best fit to the experimental parametrization
from ref. \cite{Ye:2017gyb}, achieved by variation of the two parameters $\beta$ and $m$ in
the three-quark LCwf. In particular, we obtain $m=0.28$ GeV, $\beta=0.88$ GeV, and $p=3.5$ for the power-like wavefunction, and $m=0.24$ GeV, $\beta=0.7$ GeV for the exponential wavefunction, respectively. For reference, their normalizations are given by $\mathcal{N}_{pwr}=87088.44\text{ GeV}^{-2}$ and $\mathcal{N}_{exp}=7389.46\text{ GeV}^{-2}$, respectively.
We note a satisfactory description of the electromagnetic form factors. For $Q^2$ greater than about 4~GeV$^2$ the model undershoots the
measured charge density $F_1(Q^2)$ somewhat. This could perhaps be improved in the future
by incorporating perturbative QCD corrections, which tend to increase quark transverse
momenta. Indeed, the model with the one gluon emission corrections and BFKL resummation
has been shown to reproduce fairly well the inclusive charm cross section measured at 
HERA~\cite{Dumitru:2023sjd}.

\typeout{}
\bibliography{references}

\end{document}